\documentclass{article}
\usepackage{spconf}
\usepackage{booktabs}
\usepackage{multirow}
\newcommand{\loss}{\mathcal{L}}

\newcommand{\bk}{\textless{\fontfamily{qcr}\selectfont blank}\textgreater}
\newcommand{\bh}{\boldsymbol{h}}
\newcommand{\bx}{\boldsymbol{x}}
\newcommand{\bz}{\boldsymbol{z}}
\newcommand{\by}{\boldsymbol{y}}
\newcommand{\bY}{\boldsymbol{Y}}
\newcommand{\bC}{\boldsymbol{C}}
\newcommand{\predv}{\text{Pred}^V}
\newcommand{\predb}{\text{Pred}^B}
\newcommand{\bp}{\boldsymbol{p}}
\newcommand{\bP}{\boldsymbol{P}}
\newcommand{\bX}{\boldsymbol{X}}
\newcommand{\bc}{\overset{\sim}{\boldsymbol{c}}}

\title{LongFNT: Long-form Speech Recognition with Factorized Neural Transducer}

\name{Xun Gong$^{1,2}$ \sthanks{Work done during an internship at Microsoft.}, Yu Wu$^2$, Jinyu Li$^2$, Shujie Liu$^2$, Rui Zhao$^2$, Xie Chen$^1$, Yanmin Qian$^1$}
\address{
$^1$ MoE Key Lab of Artificial Intelligence, AI Institute \\
X-LANCE Lab, Department of Computer Science and Engineering, Shanghai Jiao Tong University \\
$^2$ Microsoft
}
\begin{document}
\ninept
\maketitle
\begin{abstract}

Traditional automatic speech recognition~(ASR) systems usually focus on individual utterances, without considering long-form speech with useful historical information, which is more practical in real scenarios.
Simply attending longer transcription history for a vanilla neural transducer model shows no much gain in our preliminary experiments, since the prediction network is not a pure language model. This motivates us to leverage the factorized neural transducer structure, containing a real language model, the vocabulary predictor.
We propose the \textit{LongFNT-Text} architecture, which fuses the sentence-level long-form features directly with the output of the vocabulary predictor and then embeds token-level long-form features inside the vocabulary predictor, with a pre-trained contextual encoder RoBERTa to further boost the performance. Moreover, we propose the \textit{LongFNT} architecture by extending the long-form speech to the original speech input and achieve the best performance.
The effectiveness of our LongFNT approach is validated on LibriSpeech and GigaSpeech corpora with 19\% and 12\% relative word error rate~(WER) reduction, respectively.

\end{abstract}

\section{Introduction}
\label{sec:intro}

End-to-end~(E2E) automatic speech recognition~(ASR) models, such as connectionist temporal classification~(CTC)~\cite{graves2006ctc}, attention-based encoder-decoder~(AED)~\cite{vaswani2017attention,hori2017joint}, and recurrent neural network transducer neural transducer~(RNN-T)~\cite{graves2012sequence} are now dominating over traditional hybrid models~\cite{E2EOverview}.
A common practice is to train the E2E model with individual utterances without considering the correlation between utterances. However, real scenarios such as conversations, videos, and meetings are usually under long-form situations, which induces a significant discrepancy between training and test. To bridge the gap, long-form ASR, also known as conversational ASR, dialog-aware ASR, or large-context ASR,  is proposed to capture the relationship of transcription history.

Previous approaches explored the long-form scenario in attention-based encoder-decoder models~(AED)~\cite{hori2017joint}.
The most intuitive way is to concatenate consecutive speech or transcriptions of utterances~\cite{kim2018dialogcontext,hori2020transformerbased}. 
Hori et al.~\cite{hori2021advanced} extended their prior work to accelerate the decoding process in the streaming AED architecture.
Furthermore, consecutive long-form transcriptions can be used in recurrent neural language models~\cite{irie2019trainingb,mikolov2012contexta} or re-scoring using pretrained BERT~\cite{chiu2021crosssentence}.
A more comprehensive way is to use an auxiliary encoder to model the context information in an AED manner.
Masumura et al.~\cite{masumura2019largea,masumura2021hierarchicala} proposed a hierarchical text encoder and the distillation of large-context knowledge not limited to the current utterance, so that the long-form information can be captured while retaining the ASR model performance.
Likewise, Wei et al.~\cite{wei2022conversational,wei2022leveraging, wei2022improving} utilized a latent variational module, context-aware residual attention, and pre-trained encoders to leverage acoustic and text context.

Recently, transducer-based systems are becoming more and more popular in industry ~\cite{graves2012sequence,yeh2019transformer, tripathi2020transformera, chen2021developing,gulati2020conformer}, since it is naturally streaming with low latency, and somehow more robust than attention-based systems~\cite{chiu2019comparison, E2EOverview}. However, there is little prior work exploring long-form neural transducer models.
Narayanan et al.~\cite{narayanan2019recognizing} have done primitive explorations
by simulating long-form training and adaptation to obtain improvement using short utterances.
Schwarz et al.~\cite{schwarz2021improving} showed that the combination of input and context audio makes the network learn both speaker and environment adaptations.
 Kojima~\cite{kojima2021largecontext} explored the utilization of large context.
However, how to incorporate consecutive transcription history into the neural transducer model is still not well explored.

In this paper, we propose the novel \textit{LongFNT} to integrate long-form information into factorized neural transducer~(FNT)~\cite{chen2022factorized} architecture to solve the above challenge.
We first tried to embed long-form text into the predictor of the vanilla neural transducer, but our experiments showed that adding long-form transcriptions to the predictor part has little impact to the performance.
A possible explanation is that the prediction network in the vanilla neural transducer does not act as a pure language model (LM), which limits its capability in long-form transcription modeling. It also indicates that effective methods for AED-based models ~\cite{botros2021tied,ghodsi2020rnntransducer} cannot be extended to transducer-based ones, as they heavily rely on the LM characteristic of the decoder.
Therefore, we try to utilize long-form transcriptions using the FNT architecture, which factorizes the blank and vocabulary prediction modules so that a standalone LM can be used for vocabulary prediction.
Based on the vocabulary predictor in the FNT architecture, we propose two approaches to integrating high-level long-form features from historical transcriptions. Concretely, a context encoder is employed to yield the embedding of each token in historical transcriptions. 
Then we obtain a history embedding by an averaged pooling operation and  fuse it to the output of the vocabulary predictor, which is called \textit{sentence-level integration}.
Meanwhile, we incorporate the token embedding sequence inside the vocabulary predictor via contextual attention, which is named as \textit{token-level integration}.
We find that the two approaches complement each other to achieve better performance, referred as \textit{LongFNT-Text}.
Moreover, we employ a pre-trained text encoder, RoBERTa~\cite{roberta}, to further improve long-form ASR performance.
Finally, we embed long-form speech into the encoder to propose the \textit{LongFNT-Speech} model, and the combination of LongFNT-Text and -Speech is called \textit{LongFNT} model.
The proposed method with long-form text transcriptions achieved 17\% and 9\% relative word error rate~(WER) reduction on LibriSpeech and GigaSpeech, respectively.
By further adding long-form speech, the final improvement reached 19\% and 12\% relative WER reduction on these two tasks, respectively.

\section{Neural Transducer}
\label{sec:transducer}
\subsection{Conformer Transducer}
Conformer~\cite{gulati2020conformer} is a convolutional augmented transformer that is widely used in attention-based encoder-decoder and neural transducer architectures.
The conformer transducer~(C-T) model consists of three parts, the conformer encoder, the joint network, and the predictor network:
\begin{align}
  \bh_t &= \text{Encoder} (\bx_{\le t}), \label{eq:enc} \\
  \bz_l &= \text{Predictor} (\by_{\le l}), \\
  \bz_{t,l} &= \text{Joint} (\bh_t, \bz_l)
\end{align}
where $t, l$ are the frame and label index, respectively. The predicted probability of the neural transducer model and loss can be computed as:
\begin{align}
  P_{ASR}(\hat{y}_{l+1} | \bx_{\le t}, y_l) &= \text{softmax} (\bz_{t,l}), \\
  \loss_{transducer} &= - \log \sum_{\alpha \in \eta^{-1} (\by)} P(\alpha|\bx), \label{eq:loss_rnnt}
\end{align}
where $\eta$ is a many-to-one function from all possible transducer paths to the target $\by$.
\subsection{Modified Factorized Neural Transducer~(M-FNT) }
\begin{figure*}[ht]
  \centering
  \includegraphics[width=0.9\linewidth]{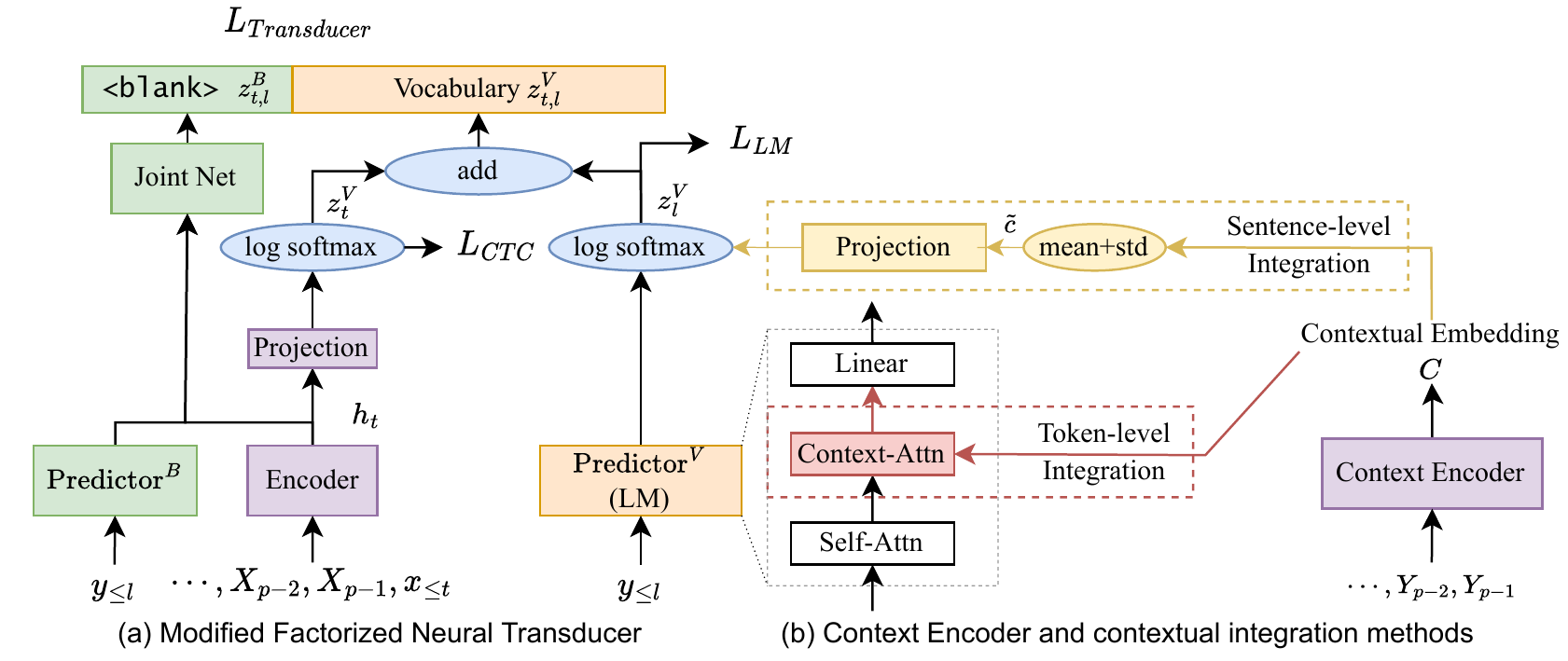}
  \caption{Architecture of modified factorized neural transducer~(M-FNT) and the long-form factorized neural transducer. Two integration methods are proposed for long-form text, and the speech encoder is extended for long-form speech.}
  \label{fig:fnt}
\end{figure*}
The factorized neural transducer model~(FNT)~\cite{chen2022factorized} aims to separately predict the blank token and vocabulary tokens, so that the vocabulary predictor fully functions as an LM.
The FNT model contains four main parts, the conformer encoder, the blank predictor~($\predb$), the joint network for \bk, and the vocabulary predictor~($\predv$, i.e. LM manner).
A main drawback of FNT is that the accuracy is slightly worse than the standard RNN-T. To tackle this issue, Zhao et al.~\cite{zhaorui} proposed some improvements for FNT, shown in Figure~\ref{fig:fnt}(a) by fusing the acoustic and label representations in the following way:
\begin{align}
  \bz_{t,l}^B &= \text{Joint} (\bh_t, \predb (\by_{\le l})), \\
  \bz_{l}^V &= \text{log\_softmax} (\predv (\by_{\le l})) \nonumber \\
       &= \log P_{LM}(\hat{y}_{l+1} | \by_{\le l}), \label{eq:predv} \\
  \bz_{t}^{V} &= \text{log\_softmax} (\text{Proj.} (\bh_t)), \nonumber \\
        &= \log P_{CTC} (\pi_t | \bx_{\le t}), \label{eq:predtv}\\
  \bz_{t,l}^V &= \bz_{t}^V + \beta \bz_{l}^V
\end{align}
where $P_{LM}$ is the predicted probability of the LM, $P_{CTC}$ is the posterior of the CTC sequence, $\beta$ is a learnable parameter.
And we can compute the posterior of the transducer model:
\begin{align}
  P_{ASR}(\hat{y}_{l+1} | \bx_{\le t}, y_l) = \text{softmax} ([\bz_{t,l}^B, \bz_{t,l}^V])
\end{align}
Finally, the modified factorized neural transducer loss is computed as
\begin{align}
  \loss = \loss_{transducer} + \lambda_{LM} \loss_{LM} + \lambda_{CTC} \loss_{CTC},
\end{align}
where $\lambda_{LM}, \lambda_{CTC}$ are hyper-parameters.

\section{LongFNT: Long-form Factorized Neural Transducer}
\label{sec:long_form}

\subsection{Long-form Context Encoder}

The long-form context encoder converts consecutive historical sentences $\{\cdots, \bY^{p-2}, \bY^{p-1}\}$ into long-form contextual embeddings $\bC$:
\begin{align}
  \bC = \text{Context-Encoder} (\cdots, \bY^{p-2}, \bY^{p-1}),
\end{align}
where the current sentences index is $p$.
The basic context encoder is jointly trained with long-form FNT which is in transformer manner.
To extract stronger features, we directly use a pre-trained RoBERTa~\cite{roberta} model as the context encoder.
The RoBERTa\footnote{https://huggingface.co/sentence-transformers/all-roberta-large-v1} is pre-trained using 160GB training text and is frozen during ASR training.

\subsection{LongFNT-Text: Long-form Context Integration}
\label{sec:long_form_text}

Here we explore different integration designs for FNT.
As shown before, FNT factorizes out the vocabulary predictor part, therefore the historical transcriptions can be injected inside the vocabulary predictor $\predv$ or after it.

\textbf{Sentence-level integration}: To integrate it in a statistical way, we first do mean and standard variance~(mean+std) of contextual embedding sequence $\bC$ to get sentence-level $\bc$, with $\predv$'s output $z^V_l$ shown in the yellow box of Figure~\ref{fig:fnt} (b), which needs an auxiliary linear layer to match the dimension:
\begin{align}
  z^{'V}_l = z^{V}_l + \text{Linear} (\bc).
\end{align}

Moreover, the contextual embedding $\bc$ can be further utilized at the linear layer $\text{Linear}^V$ of the vocabulary predictor $\predv$:
\begin{align}
  \bp &= \text{Pred-Encoder}^V (\by_{\le l}), \\
  \bP &= \text{Linear}^V \cdot \text{ReLU} (\bp), \label{eq:predv_logp} \\
  \bP' &= \text{Linear}^V \cdot \text{ReLU} (\bp \oplus \text{Projection} (\bc)),
\end{align}
where $z^{V}_l = \text{log\_softmax} (\bP \text{ or } \bP')$ to match Equation~\ref{eq:predv}, and $\oplus$ is implemented as concatenation.

\textbf{Token-level integration}: The second method is to add an auxiliary cross-attention layer inside $\predv$ transformer blocks:
\begin{align}
  \bp^i &= \text{MHA} (\bp^i, \bp^i, \bp^i), \\
  \bp^i &= \text{MHA} (\bp^i, \bC, \bC), \\
  \bp^i &= \text{FFN} (\bp^i),
\end{align}
where $i$ is the layer index, FFN is the feed-forward layer inside the transformer encoder layer and MHA is the multi-head attention layer, respectively.

Finally, the sentence-level integration and the token-level integration can be combined to achieve better utilization of contextual embedding sequence $\bC$, which is named as \textbf{LongFNT-Text}.

\subsection{LongFNT-Speech: Long-form Speech Encoder}
\label{sec:long_form_speech}

Motivated by Schwarz et al.~ \cite{schwarz2021improving}, we can propose the LongFNT-Speech encoder with extended historical speech $\{\cdots, \bX^{p-2}, \bX^{p-1}\}$ to utilize the long-form speech information:
\begin{align}
  \cdots, \bh^{p-2}, \bh^{p-1}, \bh^{p} = \text{Encoder} (\cdots, \bX^{p-2}, \bX^{p-1}, \bX^{p}), 
\end{align}
where $\bh = \bh^{p}$ matches Equation~\ref{eq:enc}.
More specifically, historical speech is only used in attention score computation, and the label representations $\bz_t^V$ is calculated by $\bh^{p}_t$ as in Equation~\ref{eq:predtv}.
Using such extension, the speech encoder receives a longer history and thus benefits training and evaluation.
Combining LongFNT-Text and LongFNT-Speech, the final proposed method is called as \textbf{LongFNT}.

\section{Experiments}
\label{sec:expr}

\subsection{Experimental Setup}
\label{sec:expr_setup}

We conduct experiments with two datasets, LibriSpeech~\cite{panayotov2015librispeech} and GigaSpeech Middle~(abbr. as GigaSpeech)~\cite{chen2021gigaspeech}. LibriSpeech has around 960 hours of audiobook speech, while GigaSpeech has around 1,000 hours of audiobook, podcasting, and YouTube audio.
The word error rate~(WER) averaged over each test set is reported.
For acoustic feature extraction, 80-dimensional mel filterbank~(Fbank) features are extracted with global level cepstral mean and variance normalization. Frame length and frame shift are 25ms and 10ms respectively.
Standard SpecAugment~\cite{park2019specaugment} is used for both datasets, respectively.
Each utterance has two frequency masks with parameter ($F=27$) and ten time masks with maximum time-mask ratio ($pS = 0.05$).
5,000 sentence pieces~\cite{kudo2018sentencepiece} are trained using LibriSpeech and GigaSpeech datasets separately.
The baseline follows the settings of FNT~\cite{chen2022factorized,zhaorui}.
The subsampling layer is a VGG2L-like network, which contains four convolution layers with the down-sampling rate of 4.
The encoder has 18 conformer layers, in which the inner size of the feed-forward layer is 1,024, and the attention dimension is 512 with 8 heads.
The $\predb$ has two LSTM layers with 1,024 hidden size and the joint dimension is set to 512.
The $\predv$ use vanilla 8 transformer layers, which has 256 attention dimension with 8 heads.
The hyper-parameter weights are fixed as $\lambda_{CTC}=0.1, \lambda_{LM}=0.5$.
The context encoder has the same shape as $\predv$ if training from scratch, and using the frozen RoBERTa model else.
As for the FNT's large LM adaptation, we use 16 transformer layers with 512 attention dimension with 8 heads.
For LibriSpeech, we use the extra text corpus, and for GigaSpeech, we use the 10,000 hours of training text data.

\subsection{Evaluation Results}
\label{sec:expr_text}

\begin{table}[ht]
  \centering
  \begin{tabular}{l|l|cccc}
    \toprule
    \multirow{2}{*}{Model} & \multirow{2}{*}{$\mathcal{M}$} & \multicolumn{2}{c}{Libri-test} & \multicolumn{2}{c}{Giga} \\
    && clean & other & dev & test \\
    \midrule
    C-T & - & 3.1 & 6.6 & 16.1 & 15.7 \\
    + long text & gt & 3.1 & 6.5 & 15.8 & 15.5 \\
    + long text & hyp & 3.1 & 6.7 & 16.1 & 15.9 \\
    + sentence-level integ. & gt & 3.1 & 6.5 & 16.0 & 15.5 \\
    + sentence-level integ. & hyp & 3.2 & 6.9 & 16.4 & 16.2 \\
    \midrule
    M-FNT       & - & 3.2 & 6.4 & 16.8 & 16.3 \\
    + long text & gt  & 3.2 & 6.3 & 16.5 & 16.2 \\
    + long text & hyp & 3.2 & 6.4 & 16.7 & 16.4 \\
    + large LM   & -    & 3.0 & 6.1 & 16.4 & 16.0 \\
    +  large LM   and long text & hyp     & 3.0 & 6.1 & 16.4 & 16.1 \\
    \midrule

    LongFNT-Text & hyp & 2.5 & 5.5 & 15.2 & 14.9 \\
    LongFNT-Speech & hyp& 2.8 & 6.0 & 15.9 & 15.7 \\
      LongFNT  & hyp & \textbf{2.4} & \textbf{5.4} & \textbf{14.8} & \textbf{14.3} \\
    \bottomrule
  \end{tabular}
  \label{tab:text}
  \caption{LongFNT performance~(WER)~(\%) on LibriSpeech test sets and GigaSpeech dev/test sets.
  $\mathcal{M}$ has two modes, `gt' denotes the historical text is obtained from ground truth, while `hyp' is from the decoded hypotheses.
  } 
\end{table}

We first explore whether the vanilla C-T model can be improved by the long transcription history, as shown in the first block of Table~\ref{tab:text}.
It indicates that the C-T model reaches little gain with extended long-form text or the sentence-level integration method. If we use it in real cases~(i.e. the previous text is decoded instead of ground truth), the performance even gets worse. Our explanation is the predictor network is not a pure language model, so it cannot utilize longer history. 

In terms of M-FNT, shown in the second block of Table~\ref{tab:text},  enlarging the text input~(`+long text') does a little help to the GigaSpeech corpus, but no more gain to LibriSpeech. When the predictor is trained with unpair text data~(+large LM), the performance of all models is increased by at least 2\%. It demonstrates that the M-FNT architecture can be benefited from a powerful language model. However, the long-form text does not benefit from large LM either, which means it is non-trivial to explore how to better leverage long-form information in the context of FNT. 

Our proposed LongFNT model achieves 19/12\% rel.~WERR compared to the M-FNT baseline and 20/9\% rel.~WERR compared to the C-T baseline. It demonstrates that LongFNT can better utilize long-form history information. Furthermore, we can observe both LongFNT-Text and LongFNT-Speech are better than baselines, indicating two types of historical information are helpful for long-form speech recognition. The LongFNT-Text is slightly better than the LongFNT-Speech model, showing the transcription history is more valuable to our model. Compared to the limited gain from the long text in C-T and M-FNT, the WER reduction from long text is relatively larger, around 10$\%$  (LongFNT v.s.  LongFNT-Speech), verifying the effectiveness of the proposed method.

\subsection{Ablation Study}

\begin{table}[ht]
  \centering
  \begin{tabular}{l|l|cccc}
    \toprule
    \multirow{2}{*}{Model} & \multirow{2}{*}{$\mathcal{M}$} & \multicolumn{2}{c}{Libri-test} & \multicolumn{2}{c}{Giga} \\
    && clean & other & dev & test \\
    \midrule
    LongFNT-Text & hyp & 2.5 & 5.5 & 15.2 & 14.9 \\
    - \textit{RoBERTa} & hyp & 2.5 & 5.6 & 15.6 & 15.2 \\
    \midrule
    - token-level integ. & hyp & 2.8 & 5.8 & 15.9 & 15.3 \\
    - - \textit{RoBERTa} & hyp & 2.9 & 5.9 & 16.0 & 15.5 \\
    - sentence-level integ. & hyp & 2.6 & 5.6 & 15.4 & 15.0 \\
    - - \textit{RoBERTa} & hyp & 2.6 & 5.7 & 15.8 & 15.4 \\
    \midrule
    \multicolumn{6}{c}{LongFNT-Text~(gt or hyp) w/o \textit{large LM} and \textit{RoBERTa}} \\ \hline
    - - token-level integ. & gt & 2.9 & 6.1 & 16.0 & 15.8 \\
    - - token-level integ. & hyp & 3.0 & 6.3 & 16.5 & 16.2 \\
    - - sentence-level integ. & gt & 2.9 & 6.0 & 15.7 & 15.4 \\
    - - sentence-level integ. & hyp & 3.0 & 6.1 & 16.0 & 15.7 \\
    \bottomrule
  \end{tabular}
    \label{tab:abl}
  \caption{The ablation study of different components in LongFNT-Text}
\end{table}

In this subsection, we validate the effectiveness of the proposed LongFNT model to explore which part is the most important.
Shown in Table~\ref{tab:abl}, we explore the efficiency of different integration methods for LongFNT-Text based on M-FNT or LongFNT-Speech architecture. The token-level integration is more important than sentence-level integration, as after removing token-level integration (i.e. only sentence-level integration), the system degrades by 12/5\% relative WER reduction in LibriSpeech and 5/3\% relative WER reduction in GigaSpeech while removing sentence-level integration only degrades $\sim$4\% and $\sim$1\% in LibriSpeech/GigaSpeech corpus.

Then, we also explore the importance of contextual encoder replacement from the train-from-scratch one to the pre-trained RoBERTa.
Results show that the importance is different from LibriSpeech and GigaSpeech datasets.
For LibriSpeech, we can see that the performance only degrades by 0.1 absolute WER reduction, and even no degradation can be found in the test-clean set for the LongFNT-Text model.
As for GigaSpeech, we can observe a consistent drop by at least 0.3$\sim$0.5 absolute WER reduction.
This interesting phenomenon is that the RoBERTa model has little impact on the LibriSpeech which may be because the influence of transcriptions is relatively smaller than the Gigaspeech dataset, and the contextual encoder has a similar ability to model the long-form textual information.

We also evaluate our models when the large pre-trained predictor network is absent. 
Detailed training data and model description can be found in the setup Section~\ref{sec:expr_setup}.
After replacing the larger in-domain LM with a trained-from-scratch $\predv$, the performance of all models decreases by at least 2\%, in the third block of Table~\ref{tab:abl}.
Furthermore, we find that long-form transcriptions are further utilized with large LM.
Strong large LM also outperforms the original one in both integration methods.

As mentioned previously in Section~\ref{sec:expr_text}, in the real scenario, ground truth~(gt) text can not be accessed, and we evaluate the above methods using decoded transcriptions~(hyp) to get real performance and explore the importance of those two types in different LongFNT-Text modes.
Shown in the 3rd block of Table~\ref{tab:abl}, models that is decoded using `(hyp)' drop the performance by 1$\sim$5\% compared to the `(gt)' one.
Results show the degradation is much higher on GigaSpeech compared with on LibriSpeech.
One possible reason is that the basic error rate influences the performance of hypo mode, a system with low WER is necessary for the  long-form speech recognition improvement. 
Moreover, we find that the sentence-level integration~(i.e. - - token-level integ.) is more sensitive to the long-form transcriptions compared to the token-level one~(i.e. - - sentence-level integ.), as it drops more 0.1$\sim$0.2 absolute WER.
And this indicates the shortcoming of statistical averaging pooling for sentence-level integration.

\subsection{The Number of Long-form Sentences}

\begin{table}[ht]
  \centering
  \begin{tabular}{l|ccc}
    \toprule
    \#Previous & 1 & 2 & 3 \\
    \midrule
    + sentence-level integ. & 16.6/16.2 & 16.4/16.1 & 16.3/16.1 \\
    ++ speech & 16.0/15.7 & 15.5/15.1 & 15.2/15.9 \\
    \bottomrule
  \end{tabular}
  \label{tab:length}
    \caption{Performance~(WER)~(\%) comparison of successive sentence counts.
  `++ speech' here means use (longFNT-Speech with sentence-level integration.
  The baseline performance is 16.8/16.3\%, which means \# Previous=0 (no historical sentence).
  }
\end{table}

From Table~\ref{tab:length}, we can find the correlation between the number of historical sentences and the recognition error rate.
As \#Previous grows, the WER of the current sentence is reduced.
However, after $\text{\#Previous} > 2$, the improvement is limited but training resources are consumed, where the most important information can be achieved in the previous two sentences.
Therefore, we select two previous consecutive sentences to learn appropriate long-form information for all experiments.

\section{Conclusion}
\label{sec:concl}

In this paper, we propose a novel long-form architecture \textbf{LongFNT} based on factorized neural transducer~(FNT) architecture.
Long-form sentence-/token-level transcription integration methods are proposed with pre-trained RoBERTa to acquire the {LongFNT-Text} model.
LongFNT-Text achieve 17\% relative WER reduction on LibriSpeech test sets and 9\% relative WER reduction on Gigaspeech corpus.
Then LongFNT-Speech is proposed to integrate long-form speech.
And {LongFNT} is obtained combining LongFNT-Text and LongFNT-Speech.
After utilizing history speech, the final system archives 19\% relative WER reduction on LibriSpeech test sets and 12\% relative WER reduction on Gigaspeech corpus.
In future work, we will explore streaming mode FNT, and more efficient training and testing methods to improve long-form speech.

\vfill\pagebreak
\bibliographystyle{IEEEbib}
\bibliography{main}
\end{document}